\documentclass[aps,pra,twocolumn,superscriptaddress]{revtex4-1}

\usepackage{graphicx,subfigure}
\usepackage{amsmath,amssymb,bm,amsthm}
\usepackage{mathtools}
\usepackage{multirow}
\usepackage{textcomp}
\usepackage{float}
\usepackage{xcolor}
\usepackage[normalem]{ulem}
\usepackage{dsfont}
\usepackage{url}
\usepackage{mathrsfs}

\newcommand{\ket}[1]{\left | #1 \right \rangle}
\newcommand{\bra}[1]{\left \langle #1 \right |}

\newcommand{\tr}{\text{Tr}}
\newcommand{\oph}{\hat{H}}

\newcommand{\opa}{\hat{a}}
\newcommand{\opad}{\hat{a}^\dagger}

\begin{document}

\title{Optimized protocols for duplex quantum transduction}

\author{Zhaoyou Wang}
\email{zhaoyou@uchicago.edu}
\author{Mengzhen Zhang}
\author{Yat Wong}
\author{Changchun Zhong}
\author{Liang Jiang}
\email{liangjiang@uchicago.edu}
\affiliation{Pritzker School of Molecular Engineering, University of Chicago, Chicago, Illinois 60637, USA}

\date{\today}

\begin{abstract}
Quantum transducers convert quantum signals through hybrid interfaces of physical platforms in quantum networks.
Modeled as quantum communication channels, performance of unidirectional quantum transduction can be measured by the quantum channel capacity.
However, characterizing performance of quantum transducers used for duplex quantum transduction where signals are converted bidirectionally remains an open question.
Here, we propose rate regions to characterize the performance of duplex quantum transduction.
Using this tool, we find that quantum transducers optimized for simultaneous duplex transduction can outperform strategies based on the standard protocol of time-shared unidirectional transduction.
Integrated over the frequency domain, we demonstrate that rate region can also characterize quantum transducers with finite bandwidth.
\end{abstract}

\maketitle

\emph{Introduction.}\textemdash
Quantum transducers convert quantum signals between physically distinct carriers, enabling quantum information exchange across multiple platforms in quantum networks~\cite{kimble2008}.
For example, a microwave-to-optical quantum transducer~\cite{lambert2020a,han2021,andrews2014,vainsencher2016,fan2018,rueda2019,jiang2020,mirhosseini2020,mckenna2020,xu2021,tu2022,shen2022} can distribute processed quantum states stored in superconducting qubits over optical fibers.
Various designs of quantum transducers have been developed, utilizing hybrid interfaces like electro-optics~\cite{tsang2010,fan2018,rueda2019,mckenna2020,xu2021}, optomechanics~\cite{aspelmeyer2014,stannigel2010,safavi-naeini2011,hill2012,palomaki2013,andrews2014,lecocq2016,vainsencher2016,jiang2020,mirhosseini2020,shen2022}, and electro-magnonics~\cite{zhang2014,tabuchi2015,shen2022}.

\begin{figure}[t]
    \centering
    \includegraphics[width=0.48\textwidth]{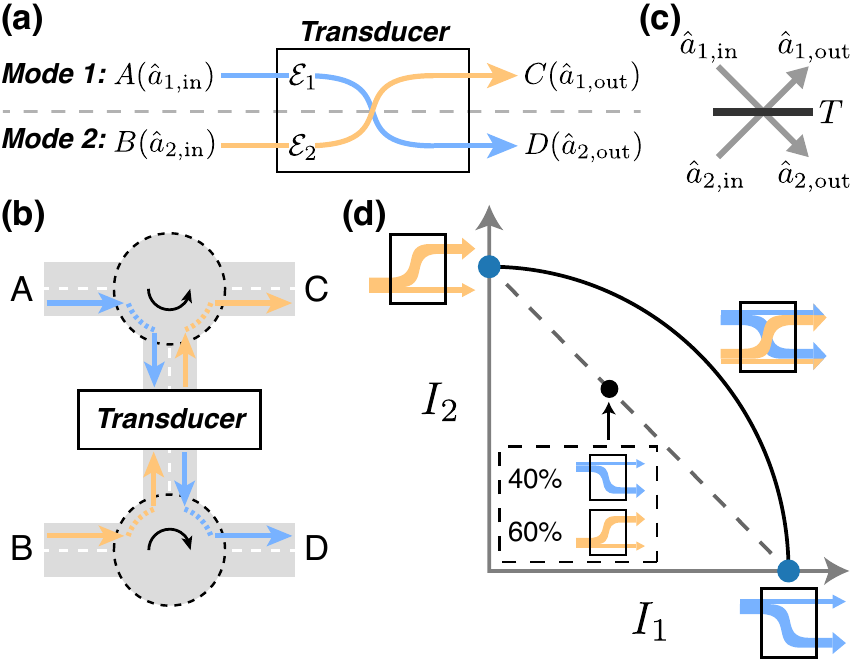}
    \caption{(a) Quantum signals converted from one mode to the other with a quantum transducer, where $\mathcal{E}_1$ and $\mathcal{E}_2$ are the unidirectional transduction channels. (b) Separating the input and output signals of each mode with circulators. (c) Beam splitter with efficiency $T$. (d) Schematic of the rate region (black line) for duplex quantum transduction. Blue dots: quantum capacity of the unidirectional transduction channels. Grey dashed line: achievable rates for the time-shared unidirectional transduction.}
    \label{fig1}
\end{figure}

As devices for quantum state transfer, quantum transducers can be abstracted as quantum channels. Most bosonic quantum transducers use red-detuned pumps~\cite{han2021,andrews2014,fan2018,rueda2019,jiang2020,mirhosseini2020,mckenna2020} to engineer a two-mode scattering process that is equivalent to a beam splitter. The input signal of one mode $\opa_{1,\text{in}} (\opa_{2,\text{in}})$ gets converted to the output signal of the other mode $\opa_{2,\text{out}} (\opa_{1,\text{out}})$, yielding two unidirectional transduction channels $\mathcal{E}_1$ and $\mathcal{E}_2$ (Fig.~\ref{fig1}(a)).
Oftentimes, only a single channel is utilized to transduce quantum signal from one mode to the other, which we refer to as \emph{unidirectional} quantum transduction.
The performance of unidirectional quantum transduction is characterized by the quantum capacity of either $\mathcal{E}_1$ or $\mathcal{E}_2$~\cite{zhang2018,wang2022b}.

By leveraging both unidirectional transduction channels, quantum signals can be converted bidirectionally which we refer to as \emph{duplex} quantum transduction~\footnote{Previous works on ``bidirectional conversion''~\cite{andrews2014,vainsencher2016,jiang2020,xu2021} measures the two unidirectional transduction channels separately on a single device, while the duplex quantum transduction we consider here is more general.}.
Duplex quantum transduction can be modeled as a quantum interference channel~\cite{fawzi2012,sen2012,das2021}, with senders $A (\opa_{1,\text{in}})$ and $B (\opa_{2,\text{in}})$ and receivers $C (\opa_{2,\text{out}})$ and $D (\opa_{1,\text{out}})$.
The senders and receivers can be distinct users by separating the input and output signals of each mode with circulators (Fig.~\ref{fig1}(b)).
For example, a lossless beam splitter with efficiency $T$ (Fig.~\ref{fig1}(c)) implements a quantum interference channel: 
\begin{equation}
    \label{eq:beam_splitter}
    \begin{split}
        \mathcal{E}_{1,A\rightarrow D}: \opa_{2,\text{out}} =& \sqrt{T} \opa_{1,\text{in}} + \sqrt{1-T} \opa_{2,\text{in}} \\
        \mathcal{E}_{2,B\rightarrow C}: \opa_{1,\text{out}} =& \sqrt{T} \opa_{2,\text{in}} - \sqrt{1-T} \opa_{1,\text{in}} .
    \end{split}
\end{equation}
One strategy for duplex quantum transduction is to alternate between using $\mathcal{E}_1$ and $\mathcal{E}_2$, while simultaneous transduction of uncorrelated input signals $\opa_{1,\text{in}}$ and $\opa_{2,\text{in}}$ may be more efficient.
However, $\mathcal{E}_1$ and $\mathcal{E}_2$ can interfere with each other when put in use simultaneously, e.g., the input signal $\opa_{1,\text{in}}$ for $\mathcal{E}_1$ acts as added noise for $\mathcal{E}_2$ in a beam splitter (Eq.~(\ref{eq:beam_splitter})).
As a result, characterizing the performance of duplex quantum transduction requires a new metric beyond the quantum capacities of the individual unidirectional transduction channels.

We propose to use achievable information rate region as the performance metric.
The achievable rates of a quantum device depend on the quantum channels it implements as well as the input signal encodings, with channel parameters (such as $T$ in Eq.~(\ref{eq:beam_splitter})) determined by the physical device parameters.
In duplex quantum transduction, both transduction channels $\mathcal{E}_1$ and $\mathcal{E}_2$ transmit quantum information at rates $I_1$ and $I_2$ respectively.
For \emph{simultaneous} duplex transduction, the pair of achievable rates $(I_1,I_2)$ depends on how we encode quantum information into the quantum signals. By varying the encodings for $\opa_{1,\text{in}}$ and $\opa_{2,\text{in}}$, we obtain a two-dimensional region of achievable information rates $\{(I_1,I_2)\}$ (Fig.~\ref{fig1}(d)).
The rate region characterizes the performance of simultaneous duplex transduction and its boundary indicates the optimized coding strategies.
Past studies have also employed rate regions or capacity regions to study the trade-off among multiple quantum channels, albeit limited to sending classical information~\cite{bennett2003,childs2006,fawzi2012,sen2012,shi2021} or distributing entanglement in qubit-based quantum networks~\cite{pant2019,shi2020,vardoyan2021,dai2022}. So far, there is no analysis investigating the achievable quantum information rate region at the hardware level, such as quantum transducers.

Furthermore, we can combine simultaneous duplex transduction with the \emph{time-sharing} strategy, where we alternate between different signal encodings and even device parameters. This leads to a new region of achievable rates that is the convex hull of the original region, and we refer to the new region as the time-sharing achievable rate region. For example, we can perform transduction in one direction with $\mathcal{E}_1$ for 40\% of the time and in the opposite direction with $\mathcal{E}_2$ for the rest 60\% of the time (black dot Fig.~\ref{fig1}(d)).
Notably, when the original region is not convex, time-sharing can offer additional performance boost for duplex quantum transduction.

In this work, we define the (time-sharing) achievable rate region and apply the tool to characterize the performance of two-mode quantum transducers. We demonstrate that a sizable portion of quantum transducers can benefit from simultaneously transducing quantum signals in both directions. We also discuss how reflectionless scattering leads to the optimal duplex quantum transduction, as well as the effect of finite bandwidth.

\emph{Rate regions of duplex quantum transduction.}\textemdash
We consider quantum transducers with a linear input-output relation
\begin{equation}
    \label{eq:linear_transducer}
    \begin{pmatrix}
        \opa_{1,\text{out}} \\
        \opa_{2,\text{out}} \\
        \vdots \\
        \opa_{n,\text{out}}
    \end{pmatrix} = S
    \begin{pmatrix}
        \opa_{1,\text{in}} \\
        \opa_{2,\text{in}} \\
        \vdots \\
        \opa_{n,\text{in}}
    \end{pmatrix} ,
\end{equation}
where the scattering matrix $S$ is unitary and depends on the device parameters of the transducer~\cite{han2021}. We choose ports 1 and 2 as the signal ports, and $\opa_{n>2,\text{in}}$ are the injected vacuum noise from the internal loss channels.
The two transduction channels are
\begin{equation}
    \label{eq:two_channels}
    \begin{split}
        \mathcal{E}_{1,A\rightarrow D} : \opa_{2,\text{out}} =& S_{21} \opa_{1,\text{in}} + S_{22} \opa_{2,\text{in}} + \sum_{n>2} S_{2n} \opa_{n,\text{in}} \\
        \mathcal{E}_{2,B\rightarrow C}: \opa_{1,\text{out}} =& S_{12} \opa_{2,\text{in}} + S_{11} \opa_{1,\text{in}} + \sum_{n>2} S_{1n} \opa_{n,\text{in}} .
    \end{split}
\end{equation}
Intuitively, the transmission coefficients $S_{12}$ and $S_{21}$ determine the transduction efficiency, while the reflection coefficients $S_{11}$ and $S_{22}$ lead to the interference between $\mathcal{E}_1$ and $\mathcal{E}_2$.

Here we define the achievable information rates for simultaneous duplex transduction.
For a quantum channel $\mathcal{E}: \mathcal{L}(\mathcal{H}) \rightarrow \mathcal{L}(\mathcal{H})$, the achievable rate of quantum information with an input state $\hat{\rho}$ is measured by the coherent information $I(\mathcal{E}, \hat{\rho})$~\cite{wilde2013}. Let $\ket{\psi} \in \mathcal{H} \otimes \mathcal{H}'$ be a purification of $\hat{\rho}$, we have
\begin{equation}
    \label{eq:coherent_information}
    I(\mathcal{E}, \hat{\rho}) \equiv H(\mathcal{E} (\hat{\rho})) - H ((\mathcal{E}\otimes\mathcal{I}') (\ket{\psi} \bra{\psi})), 
\end{equation}
where $H(\hat{\rho})$ is the von Neumann entropy of $\hat{\rho}$ and $\mathcal{I}'$ is the identity map on $\mathcal{H}'$.
For $I(\mathcal{E}, \hat{\rho})<0$, the achievable rate is 0.
Generalizing to a quantum interference channel $\mathcal{E}_{(A,B)\rightarrow (C,D)}: \mathcal{L}(\mathcal{H}_1 \otimes \mathcal{H}_2) \rightarrow \mathcal{L}(\mathcal{H}_1 \otimes \mathcal{H}_2)$, the simultaneously achievable information rates $(I_1,I_2)$ with uncorrelated input state $\hat{\rho}_1 \otimes \hat{\rho}_2$ are
\begin{equation}
    \label{eq:general_rates}
    \begin{split}
        I_1 \left(\mathcal{E}, \hat{\rho}_1 \otimes \hat{\rho}_2 \right) \equiv & I \left(\mathcal{E}_1, \hat{\rho}_1 \right) \qquad \mathcal{E}_1 (\cdot) = \tr_1 \mathcal{E} (\cdot, \hat{\rho}_2) \\
        I_2 \left(\mathcal{E}, \hat{\rho}_1 \otimes \hat{\rho}_2 \right) \equiv & I \left(\mathcal{E}_2, \hat{\rho}_2 \right) \qquad \mathcal{E}_2 (\cdot) = \tr_2 \mathcal{E} (\hat{\rho}_1, \cdot) .
    \end{split}
\end{equation}

\begin{figure}
    \centering
    \includegraphics[width=0.48\textwidth]{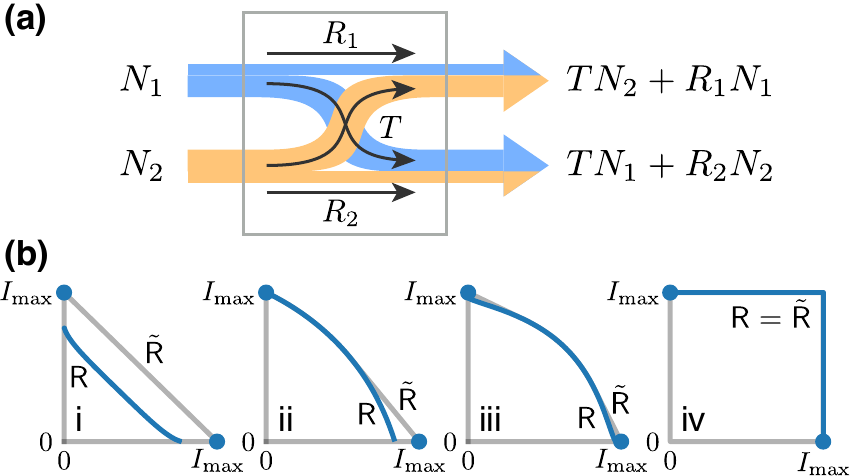}
    \caption{(a) Duplex transduction with thermal input states. (b) The rate regions (blue dot and lines) and time-sharing rate regions (grey lines) at $T=0.9$ for different reflection coefficients $(R_1,R_2)=(0.03,0.03),(0.03,0),(0.003,0.003),(0,0)$.}
     \label{fig2}
\end{figure}

Given the challenges in determining the quantum capacity for lossy channels with added noise~\cite{schumacher1996,lloyd1997,devetak2005a,weedbrook2012}, we focus on the rate region achievable with thermal input states as a lower bound.
When the input signals of $\mathcal{E}_1$ and $\mathcal{E}_2$ are thermal states with average photon number $N_1$ and $N_2$, the outputs are also thermal states with photon number $N_1' = T N_1+R_2 N_2$ and $N_2' = T N_2 + R_1 N_1$ (Fig.~\ref{fig2}(a)).
Here $R_i=|S_{ii}|^2$ is the power reflection coefficient from port $i$, $T_{ij}=|S_{ij}|^2$ is the power transmission coefficient from port $j$ to port $i$, and we assume $T_{12}=T_{21} \equiv T$.
For finite $R_1$ and $R_2$, the reflected signal from one channel adds thermal noise to the other channel, which leads to the trade-off between $I_1$ and $I_2$ for simultaneous duplex transduction.

The achievable rates $(I_1,I_2)$ for Eq.~(\ref{eq:two_channels}) with thermal input states $(N_1,N_2)$ are (see Appendix~\ref{SI_A})
\begin{equation}
    \label{eq:rates}
    \begin{split}
        & I_k (T,R_1,R_2,N_1,N_2) = h(N_k') \\
        & - h\left( \frac{D_k + N_k' - N_k - 1}{2} \right) - h\left( \frac{D_k - N_k' + N_k - 1}{2} \right) ,
    \end{split}
\end{equation}
where $ k=1,2$,
\begin{equation}
    h(x) = (x+1) \log_2 (x+1) - x\log_2 (x) ,
\end{equation}
and
\begin{equation}
    D_k = \sqrt{(N_k+N_k'+1)^2 - 4T N_k (N_k+1)} .
\end{equation}
The rate region $\mathsf{R} \equiv \{(I_1,I_2) | \forall (N_1,N_2)\}$ only depends on channel parameters $(T,R_1,R_2)$. We could combine simultaneous duplex transduction with the time-sharing protocol, and the resulting time-sharing rate region is the convex hull $\tilde{\mathsf{R}} = \text{Conv}(\mathsf{R})$.
Additionally, numerical evidences suggest that thermal encodings are likely optimal among general Gaussian encodings (see Appendix~\ref{SI_A1}).
The rate regions can be calculated similarly when the environment injects thermal noise rather than vacuum noise via the internal loss channels $\opa_{n>2,\text{in}}$ (see Appendix~\ref{SI_A2}).

The rate region $\mathsf{R}$ can be determined from its boundary $\partial \mathsf{R}$. For the special cases of unidirectional quantum transduction with $(N_1,N_2)=(\infty,0)$ and $(0,\infty)$, we achieve information rates $(I_{\text{max}},0)$ and $(0,I_{\text{max}})$ on $\partial \mathsf{R}$ (Fig.~\ref{fig2}(b), blue dots). Here $I_{\text{max}} = \max \{\log_2 (T/(1-T)),0\} $ is the quantum capacity of the pure-loss channel~\cite{holevo2001}.
For $I_1>0$ and $I_2>0$, $\mathsf{R}$ corresponds to a continuous mapping $(N_1,N_2) \rightarrow (I_1,I_2)$ and $\partial \mathsf{R}$ can be solved numerically with the low-rank Jacobian condition $\det (J) = 0$, where $J$ is the $2\times 2$ Jacobian matrix.
In Fig.~\ref{fig2}(b), we plot the rate regions $\mathsf{R}$ (blue lines and dots) and $\tilde{\mathsf{R}}$ (grey lines) for different reflection coefficients $(R_1,R_2)$. We choose $T=0.9$ with $I_{\text{max}} \approx 3.17$.

Finite reflection $R_k$ results in a noticeable discontinuity of the boundary $\partial \mathsf{R}$ at the $I_k$ axis (Fig.~\ref{fig2}(b)i-iii). This can be explained from the upper bound on the thermal-loss capacity~\cite{rosati2018,sharma2018,noh2019}.
Assuming $R_1 > 0$, the channel $\mathcal{E}_{2,B\rightarrow C}$ is a thermal loss channel with noise photon $\bar{N}=R_1N_1/(1-T)$.
From the upper bound~\cite{rosati2018,sharma2018,noh2019}
\begin{equation}
    \label{eq:upper_bound}
    I_2 \leq \max \left\{ \log_2 \left[ \frac{T-(1-T) \bar{N}}{(1-T)(\bar{N}+1)} \right], 0 \right\} ,
\end{equation}
we must have $N_1 < (2T-1)/2R_1$ to achieve a positive information rate $I_2>0$. On the other hand, when $I_2=0$ the quantum capacity $I_1=I_{\text{max}}$ is achieved at $N_1 \rightarrow \infty$, which leads to the discontinuity at $I_2 = 0$.

If one side is reflectionless with $R_k=0$, the discontinuity of $\partial \mathsf{R}$ vanishes at the $I_k$ axis (Fig.~\ref{fig2}(b)ii $I_2$ axis). If both sides are reflectionless, there is no interference between the two transduction channels and the maximal square region can be achieved (Fig.~\ref{fig2}(b)iv). Therefore it is possible to outperform the time-shared unidirectional transduction (Fig.~\ref{fig1}(d) grey dashed line) with the simultaneous duplex transduction, as long as the reflection coefficients are small (Fig.~\ref{fig2}(b)ii-iv).

So far we have only considered direct transduction without adaptive control~\cite{zhang2018} or shared entanglement~\cite{zhong2020a,zhong2020,wu2021}.
In Appendix~\ref{SI_B}, we briefly discuss duplex quantum transduction assisted with local operations and classical communication (LOCC), along with the scenario where the senders are the same as the receivers which allows the interference-based techniques~\cite{lau2019,zhang2022a}.

\emph{Optimized transduction protocols.}\textemdash
Here we apply the tool of rate regions to analyze a physical transducer model.
The channel parameters $(T,R_1,R_2)$, and thus the achievable rates, depend on the device parameters of the transducer. We therefore generalize the achievable rate regions to include not only different signal encodings but also different device parameters.
The boundary of the resulting rate region leads to optimized signal encodings and device parameters for the transducer.

\begin{figure}
    \centering
    \includegraphics[width=0.48\textwidth]{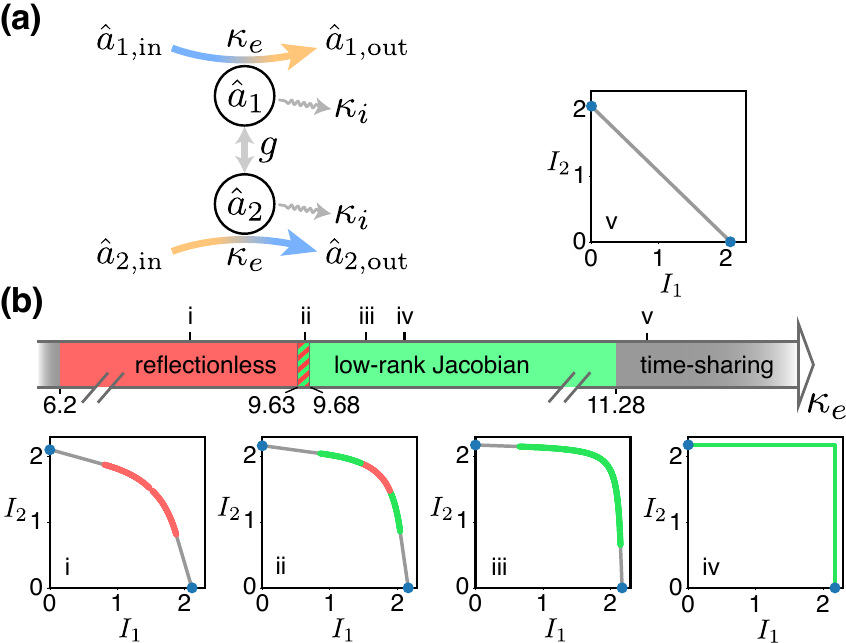}
    \caption{(a) Schematic of a physical transducer model. (b) Optimized protocols achieving the boundary $\partial \tilde{\mathsf{R}}$ for the time-shared duplex transduction at different $\kappa_e$. Here i-v correspond to $\kappa_e=9.2,9.66,9.9,10.05,11.4$ respectively.}
     \label{fig3}
\end{figure}

We consider a transducer model for frequency conversion between two bosonic modes $\opa_1$ and $\opa_2$ (Fig.~\ref{fig3} (a)). The lab frame Hamiltonian is
\begin{equation}
    \oph = \omega_1 \opad_1 \opa_1 + \omega_2 \opad_2 \opa_2 + g \left( \opad_1 \opa_2 e^{i\omega_p t} + \opa_1 \opad_2 e^{-i\omega_p t} \right) ,
\end{equation}
where $\omega_k$ are the mode frequencies, $\omega_p$ is the pump frequency and $g$ is the interaction rate.
An input signal at frequency $\omega$ in mode 1 gets converted to an output signal at frequency $\omega+\omega_p$ in mode 2, and vice versa.
The Hamiltonian in the rotating frame of the signal is
\begin{equation}
    \label{eq:transducer_Hamiltonian}
    \oph = \Delta_1 \opad_1 \opa_1 + \Delta_2 \opad_2 \opa_2 + g \left( \opad_1 \opa_2 + \opa_1 \opad_2 \right) ,
\end{equation}
where $\Delta_1 = \omega_1 - \omega$ and $\Delta_2 = \omega_2 - \omega_p - \omega$.

Assuming mode $k$ has external (internal) loss rate $\kappa_{k,e(i)},k=1,2$, the scattering matrix only depends on the ratios $\kappa_{k,e}/\kappa_{k,i}$ (see Appendix~\ref{SI_D}).
In practice, the loss rates of two modes may differ by orders of magnitude, but the ratios $\kappa_{1,e}/\kappa_{1,i}$ and $\kappa_{2,e}/\kappa_{2,i}$ are often close~\cite{fan2018,mckenna2020,xu2021}.
Therefore, we assume symmetric loss rates $\kappa_{1,e(i)}=\kappa_{2,e(i)} \equiv \kappa_{e(i)}$ for simplicity and the more general case is discussed in Appendix~\ref{SI_D}.
The input-output relation is given by
\begin{equation}
    \begin{pmatrix}
        \opa_{1,\text{out}} \\
        \opa_{2,\text{out}} \\
        \tilde{a}_{1,\text{out}} \\
        \tilde{a}_{2,\text{out}} 
    \end{pmatrix} = S
    \begin{pmatrix}
        \opa_{1,\text{in}} \\
        \opa_{2,\text{in}} \\
        \tilde{a}_{1,\text{in}} \\
        \tilde{a}_{2,\text{in}} 
    \end{pmatrix} ,
\end{equation}
where $\tilde{a}_{k,\text{in(out)}}$ are the internal loss channels, and
\begin{equation}
    \label{eq:S_matrix}
    S = 
    \begin{pmatrix}
        I+\kappa_e M & \sqrt{\kappa_e \kappa_i} M \\
        \sqrt{\kappa_e \kappa_i} M & I+\kappa_i M 
    \end{pmatrix} .
\end{equation}
Here $I$ is the $2\times 2$ identity matrix, and
\begin{equation}
    M = - \left( iG + \frac{\kappa_e+\kappa_i}{2}I \right)^{-1}, \quad
    G = 
    \begin{pmatrix}
        \Delta_1 & g \\
        g & \Delta_2
    \end{pmatrix} .
\end{equation}

We focus on optimizing the detunings $(\Delta_1,\Delta_2)$ of the transducer, while keeping other relevant parameters $(g, \kappa_e, \kappa_i)$ fixed.
Besides the signal encodings $(N_1,N_2)$, the achievable rates $(I_1,I_2)$ also depends on the device parameters $(\Delta_1,\Delta_2,g,\kappa_e,\kappa_i)$.
We therefore define the rate region as $\mathsf{R} \equiv \{(I_1,I_2) | \forall (N_1,N_2,\Delta_1,\Delta_2)\}$. The optimized signal encodings $(N_1,N_2)$ and detunings $(\Delta_1,\Delta_2)$ can be obtained from the boundary $\partial \tilde{\mathsf{R}}$ of the time-sharing rate region $\tilde{\mathsf{R}}$.

The boundary of the rate region can be determined by exploring several possible solutions. On the $I_1$ and $I_2$ axes, the quantum capacity of unidirectional quantum transduction increases with the transmission coefficient. Therefore we choose $(\Delta_1,\Delta_2)$ that leads to the highest transmission rate $T$ to achieve the information rates $(I_{\text{max}},0)$ and $(0,I_{\text{max}})$ on $\partial \mathsf{R}$.
For $I_1>0$ and $I_2>0$, $\mathsf{R}$ corresponds to a continuous mapping $(N_1,N_2,\Delta_1,\Delta_2) \rightarrow (I_1,I_2)$. The boundary $\partial \mathsf{R}$ as extreme values of the mapping can be obtained by comparing two possible solutions.
One solution is from the low-rank Jacobian condition $\text{rank}(J)<2$, where $J$ is the $2\times 4$ Jacobian matrix.
The other solution is from the reflectionless condition with $R_k=0$ and $N_k \rightarrow \infty$ where $k=1$ or $2$.
We consider this solution separately since the Jacobian matrix may be undefined under the limit of $N_k \rightarrow \infty$.
The reflectionless solution can be calculated analytically. For example, $R_1=0$ requires
\begin{equation}
    \Delta_1 = \frac{\kappa_{e} - \kappa_{i}}{\kappa_{e} + \kappa_{i}} \Delta_2, \qquad \Delta_2 = \sqrt{ \frac{(\kappa_{e} + \kappa_{i}) (4g^2 - \kappa_{e}^2 + \kappa_{i}^2)}{4 (\kappa_{e} - \kappa_{i})} },
\end{equation}
which leads to the achievable rates at $N_1 \rightarrow \infty$
\begin{equation}
    \begin{split}
        I_1(N_2) =& \log_2 \frac{T}{1-T} - h \left( R_2 N_2 \frac{1+T}{2(1-T)} \right) \\
        I_2(N_2) =& h(TN_2) - h((1-T)N_2) .
    \end{split}
\end{equation}
In practice, we expect an approximate reflectionless solution with $R_k \approx 0$ and finite $N_k$, due to input power constraints and uncertainties in controlling the reflection coefficients.

We calculate the time-sharing rate region $\tilde{\mathsf{R}}(\kappa_e)$ for several choices of $\kappa_e$ at $g=5$ and $\kappa_i=1$ (Fig.~\ref{fig3}(b)).
The boundary $\partial \tilde{\mathsf{R}}$ may be composed of one or more types of the protocols: reflectionless (red), low-rank Jacobian (green) and time-sharing (grey).
For example, for $9.63 < \kappa_e < 9.68$ the boundary contains all three types of protocols (Fig.~\ref{fig3}(b)ii) while for $\kappa_e > 11.28$ or $\kappa_e < 6.2$ the optimized protocols is time-shared unidirectional transduction.
It is also worth mentioning that for $6.2 < \kappa_e < 11.28$ the transducers benefit from the simultaneous duplex transduction.

If both $\kappa_e$ and detunings are tunable, it can be proved that the highest transmission rate $T$ occurs when $R_1=R_2=0$ (see Appendix~\ref{SI_C}). Therefore the optimal duplex quantum transduction is achieved with the two-side reflectionless condition at $\kappa_e=\sqrt{4g^2 + \kappa_i^2}$ and $\Delta_1=\Delta_2=0$ (Fig.~\ref{fig3}(b)iv).
In other words, $\tilde{\mathsf{R}}(\sqrt{4g^2 + \kappa_i^2})$ is the largest possible region in the sense that $\tilde{\mathsf{R}}(\kappa_e) \subseteq \tilde{\mathsf{R}}(\sqrt{4g^2 + \kappa_i^2})$ for any $\kappa_e$.

\emph{Frequency-integrated rate region.}\textemdash
Quantum transducer usually has a finite conversion bandwidth, which determines the range of signal frequencies that can be converted efficiently~\cite{zeuthen2020}.
Larger bandwidth enables higher operation speed of the transducer, and is preferable in presence of decoherence.
Within the bandwidth, quantum signals at multiple frequencies can be transduced independently with frequency-dependent conversion efficiencies.

\begin{figure}
    \centering
    \includegraphics[width=0.48\textwidth]{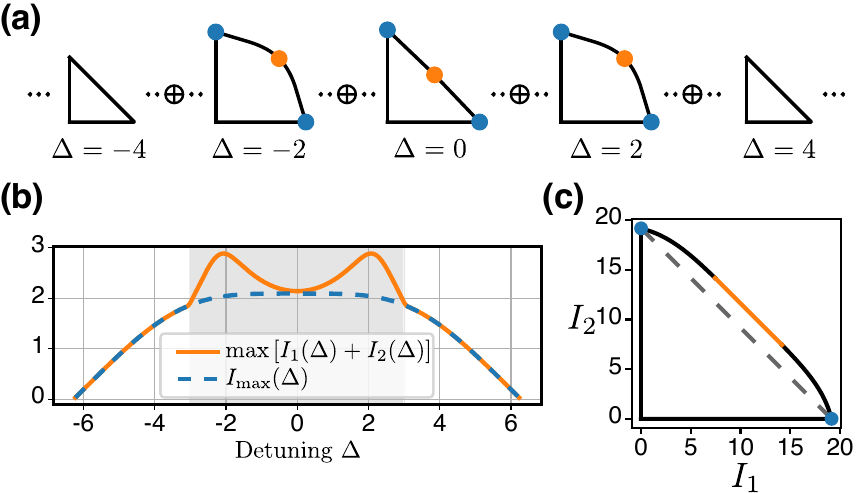}
    \caption{(a) The time-sharing rate regions $\tilde{\mathsf{R}}(\Delta)$ for different signal frequencies. Here $g=5,\kappa_e=9,\kappa_i=1$ are fixed. (b) Comparing the maximal $I_1(\Delta)+I_2(\Delta)$ with the quantum capacity $I_{\text{max}}(\Delta)$. (c) Frequency-integrated rate region $\tilde{\mathsf{R}}_{\text{tot}}$ over all regions $\tilde{\mathsf{R}} (\Delta)$ in (a).}
     \label{fig4}
\end{figure}

We can perform duplex quantum transduction in parallel for various signal frequencies, and the frequency-dependent scattering matrices result in distinct achievable rate regions that vary with frequency.
Let $\omega_2=\omega_p+\omega_1$ for the transducer model Eq.~(\ref{eq:transducer_Hamiltonian}), the signal detuning in the rotating frame becomes $\Delta_1 = \Delta_2 \equiv \Delta$, and the frequency-dependent rate region is $\mathsf{R} (\Delta) \equiv \{ (I_1(\Delta), I_2 (\Delta)) | \forall (N_1,N_2) \}$.
We plot the time-sharing rate regions $\tilde{\mathsf{R}} (\Delta)$ for multiple $\Delta$ (Fig.~\ref{fig4}(a)), and compare $\max [I_1 (\Delta) + I_2(\Delta) ]$ with the quantum capacity $I_{\text{max}}(\Delta)$ (Fig.~\ref{fig4}(b)). For $\Delta$ within the grey shaded region, simultaneous duplex transduction is advantageous, while outside this region time-shared unidirectional transduction is the optimal protocol.

To obtain the total achievable rate region, we sum the contributions from the individual rate regions at each signal frequency (Fig.~\ref{fig4}(a)).
The frequency-integrated rate region for time-shared duplex transduction is defined as
\begin{equation}
    \begin{split}
        \tilde{\mathsf{R}}_{\text{tot}} \equiv & \int^{\oplus} \tilde{\mathsf{R}}(\omega) \text{d}\omega \\
        \approx & \delta (\cdots \oplus \tilde{\mathsf{R}}(\omega_{j-1}) \oplus \tilde{\mathsf{R}}(\omega_j) \oplus \tilde{\mathsf{R}}(\omega_{j+1}) \oplus \cdots) ,
    \end{split}
\end{equation}
where $\{\omega_j\}$ is a set of evenly spaced frequencies with a frequency spacing $\delta$, and $\mathsf{A} \oplus \mathsf{B} \equiv \{ \bm{a}+\bm{b} | \bm{a} \in \mathsf{A}, \bm{b} \in \mathsf{B} \}$ is the Minkowski sum~\cite{deberg2008} of two sets $\mathsf{A}$ and $\mathsf{B}$.
For general sets the complexity of Minkowski sum is $O(|A||B|)$, while for convex sets $\mathsf{A}$ and $\mathsf{B}$ in $\mathbb{R}^2$ the complexity is $O(|\partial A| + |\partial B|)$~\cite{deberg2008}. Therefore numerical evaluation of $\tilde{\mathsf{R}}_{\text{tot}}$ is efficient since $\tilde{\mathsf{R}}$ is convex in $\mathbb{R}^2$.

We calculate the Minkowski sum of all rate regions $\tilde{\mathsf{R}} (\Delta)$ in Fig.~\ref{fig4}(a), and the resulting frequency-integrated rate region $\tilde{\mathsf{R}}_{\text{tot}}$ is shown in Fig.~\ref{fig4}(c). The boundary $\partial \tilde{\mathsf{R}}_{\text{tot}}$ can be achieved with frequency dependent protocols.
For example, to realize the orange part of $\partial \tilde{\mathsf{R}}_{\text{tot}}$ with a slope of -1, we choose the simultaneous duplex transduction protocol that maximizes $I_1 (\Delta) + I_2(\Delta)$ (Fig.~\ref{fig4}(a) orange dots) for signal detuning $\Delta$ within the grey shaded region in Fig.~\ref{fig4}(b). For other $\Delta$, we perform the time-shared unidirectional transduction.
Benefiting from simultaneous duplex transduction, the frequency-integrated rate region outperforms the time-shared unidirectional transduction (Fig.~\ref{fig4}(c) grey dashed line).

\emph{Discussion.}\textemdash
We proposed the (time-sharing) rate region to quantify the performance of duplex quantum transduction, and studied optimized protocols for a two-mode quantum transducer.
Unlike unidirectional quantum transduction, duplex quantum transduction is influenced by the reflection coefficients, and we explored how the reflectionless condition can be related to the optimal duplex quantum transduction.
Furthermore, we incorporated the finite bandwidth of the transducer and introduced the frequency-integrated rate region.
In future works, it would be interesting to consider non-Gaussian encodings (see Appendix~\ref{SI_E}), as well as other approaches to quantum transduction such as adaptive control~\cite{zhang2018}, shared entanglement~\cite{zhong2020a,zhong2020,wu2021} and interference-based methods~\cite{lau2019,zhang2022a}.
Our method can also be extended to analyze the performance of multiplex quantum hardware with more than two quantum channels, such as characterizing the performance of a 3-port quantum circulator with a 3-dimensional rate region.
Exploring alternative performance metrics~\cite{siddiqui2022} for duplex quantum transduction may provide further insights.

\begin{acknowledgments}
We thank Amir Safavi-Naeini, Cheng Guo, Chiao-Hsuan Wang, Mark M. Wilde and Siddhartha Das for helpful discussions. We acknowledge support from the ARO(W911NF-23-1-0077), ARO MURI (W911NF-21-1-0325), AFOSR MURI (FA9550-19-1-0399, FA9550-21-1-0209), AFRL (FA8649-21-P-0781), NSF (OMA-1936118, ERC-1941583, OMA-2137642), NTT Research, and the Packard Foundation (2020-71479). L.J. acknowledges the support from the Marshall and Arlene Bennett Family Research Program. This material is based upon work supported by the U.S. Department of Energy, Office of Science, National Quantum Information Science Research Centers.
\end{acknowledgments}

\appendix

\section{Achievable rates of linear transducers}
\label{SI_A}
We provide details for the achievable rates Eq.~(\ref{eq:rates}) and discuss about the capacity region as well as rate region with general Gaussian states.
The derivations are based on properties of Gaussian states and channels~\cite{weedbrook2012,noh2019}, and we follow the convention in Ref.~\cite{noh2019}.

The transducer defined in Eq.~(\ref{eq:linear_transducer}) is a passive linear system, i.e., linear system without squeezing. Since $S$ is unitary, Eq.~(\ref{eq:linear_transducer}) is a Gaussian unitary transformation described by a $2n\times 2n$ symplectic matrix
\begin{equation}
    \bm{\mathrm{S}} = 
    \begin{pmatrix}
        \bm{\mathrm{S}}_{11} & \bm{\mathrm{S}}_{12} & \cdots \\
        \bm{\mathrm{S}}_{21} & \bm{\mathrm{S}}_{22} & \cdots \\
        \vdots & \vdots & \ddots
    \end{pmatrix} 
\end{equation}
with
\begin{equation}
    \bm{\mathrm{S}}_{ij} = 
    \begin{pmatrix}
        \text{Re} \{S_{ij}\} & -\text{Im} \{S_{ij}\} \\
        \text{Im} \{S_{ij}\} & \text{Re} \{S_{ij}\}
    \end{pmatrix}
    = |S_{ij}| \bm{\mathrm{R}} (\theta_{ij}) ,
\end{equation}
where $\theta_{ij}=\arg (S_{ij})$ and we have defined the phase rotation matrix
\begin{equation}
    \bm{\mathrm{R}} (\theta) = 
    \begin{pmatrix}
        \cos \theta & -\sin \theta \\
        \sin \theta & \cos \theta
    \end{pmatrix} .
\end{equation}

The coherent information Eq.~(\ref{eq:coherent_information}) can be calculated analytically for Gaussian input states. We choose a $n$-mode Gaussian state with covariance matrix $\bm{\mathrm{V}} = \text{diag}(\bm{\mathrm{V}}_1, \bm{\mathrm{V}}_2, \cdots, \bm{\mathrm{V}}_n)$, where $\bm{\mathrm{V}}_k$ is the covariance matrix for mode $k$.
For quantum channel $\mathcal{E}_{1,A\rightarrow D}$, we can purify the input state of mode 1 with a reference system. The joint covariance matrix of the reference system and mode 1 is
\begin{equation}
    \begin{pmatrix}
        \bm{\mathrm{V}}_1^{\oplus} & \bm{\mathrm{C}}_1^T \\
        \bm{\mathrm{C}}_1 & \bm{\mathrm{V}}_1
    \end{pmatrix} ,
\end{equation}
where
\begin{equation}
    \begin{split}
        \bm{\mathrm{V}}_1^\oplus=&\left(N_1+\frac{1}{2} \right)\bm{\mathrm{I}} \\
        \bm{\mathrm{C}}_1=& \sqrt{N_1 (N_1+1)} \bm{\mathrm{R}}(\theta_1) \bm{\mathrm{S}}(r_1) \bm{\mathrm{Z}} 
    \end{split}
\end{equation}
and the decomposition of general one-mode Gaussian states is
\begin{equation}
    \label{eq:gaussian_decomposition}
    \bm{\mathrm{V}}_1= \left( N_1+\frac{1}{2} \right) \bm{\mathrm{R}}(\theta_1) \bm{\mathrm{S}}(2 r_1) \bm{\mathrm{R}}^T(\theta_1) .
\end{equation}
Here $r_1$ is the squeezing parameter and $\theta_1$ is the angle of the squeezing axis. We have defined $\bm{\mathrm{I}} = \text{diag} (1,1), \bm{\mathrm{Z}} = \text{diag} (1,-1)$ and the single mode squeezing transformation $\bm{\mathrm{S}} (r) = \text{diag} (e^{-r},e^r)$.

The coherent information for channel $\mathcal{E}_{1,A\rightarrow D}$ is
\begin{equation}
    \label{eq:general_Ic}
    I_{1,A\rightarrow D} = H(\bm{\mathrm{V}}_2') - H(\bm{\mathrm{V}}_2'')
\end{equation}
where $H(\bm{\mathrm{V}})$ is the von Neumann entropy of a Gaussian state with covariance matrix $\bm{\mathrm{V}}$, and
\begin{equation}
    \begin{split}
        \bm{\mathrm{V}}_2'=&\sum_{k=1}^n \bm{\mathrm{S}}_{2k} \bm{\mathrm{V}}_k \bm{\mathrm{S}}_{2k}^T \\
        \bm{\mathrm{V}}_{2}''=&
        \begin{pmatrix}
            \bm{\mathrm{V}}_1^{\oplus} & \bm{\mathrm{C}}_1^T \bm{\mathrm{S}}_{21}^T \\
            \bm{\mathrm{S}}_{21} \bm{\mathrm{C}}_1 & \bm{\mathrm{V}}_2'
        \end{pmatrix} .
    \end{split}
\end{equation}
Here $\bm{\mathrm{V}}_2'$ is the output state of mode 2 and $\bm{\mathrm{V}}_{2}''$ is the joint output state of the reference system and mode 2.

We assume vacuum input states for modes $n > 2$ with $\bm{\mathrm{V}}_{n>2}= \frac{1}{2} \bm{\mathrm{I}}$. In this case,
\begin{equation}
    \begin{split}
        \bm{\mathrm{V}}_2'=&\sum_{k=1}^n \bm{\mathrm{S}}_{2k} \bm{\mathrm{V}}_k \bm{\mathrm{S}}_{2k}^T \\
        =& \bm{\mathrm{S}}_{21} \bm{\mathrm{V}}_1 \bm{\mathrm{S}}_{21}^T + \bm{\mathrm{S}}_{22} \bm{\mathrm{V}}_2 \bm{\mathrm{S}}_{22}^T + \frac{1}{2} (1-|S_{21}|^2-|S_{22}|^2) \bm{\mathrm{I}}
    \end{split}
\end{equation}
where we have used the fact that $S$ is unitary.
Therefore $I_1$ only depends on $S_{21},S_{22}$ and $\bm{\mathrm{V}}_1,\bm{\mathrm{V}}_2$. Similarly, $(I_1,I_2)$ only depends on $S_{11},S_{12},S_{21},S_{22}$ and $\bm{\mathrm{V}}_1,\bm{\mathrm{V}}_2$.
If the input states to mode 1 and 2 are thermal states instead of general Gaussian states, Eq.~(\ref{eq:general_Ic}) reduces to Eq.~(\ref{eq:rates}) and the rate region $\mathsf{R}$ only depends on $(T,R_1,R_2)$.

\begin{figure}
    \centering
    \includegraphics[width=0.48\textwidth]{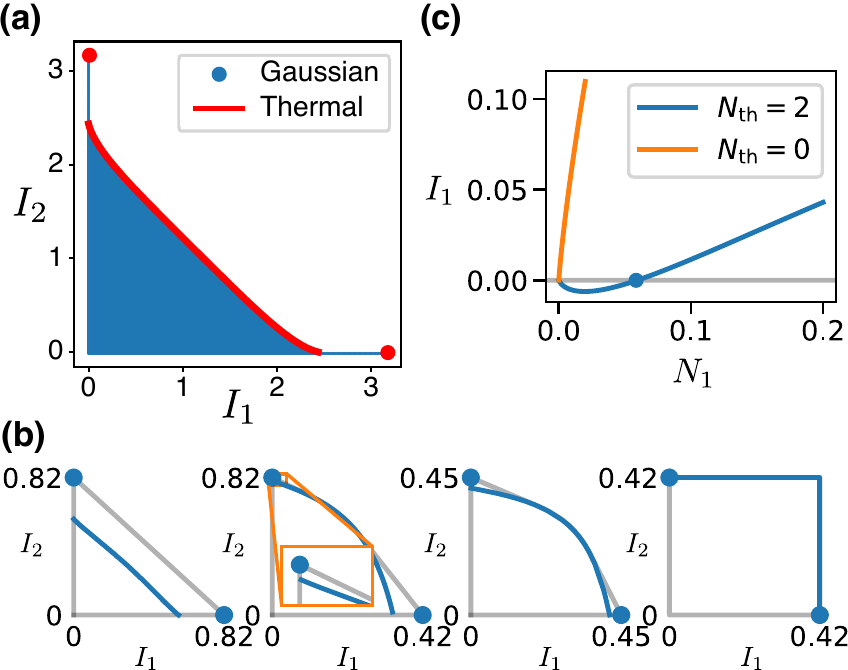}
    \caption{(a) Blue region: $\mathsf{R}_{\text{gaussian}}$, red line and red dots: $\partial \mathsf{R}_{\text{thermal}}$. Numerically we have $\mathsf{R}_{\text{thermal}} = \mathsf{R}_{\text{gaussian}}$. Here $\mathsf{R}_{\text{gaussian}}$ is calculated by sweeping $N_1,N_2$ from $10^{-3}$ to $10^3$, $r_1,r_2$ from 0 to 2, and $\Delta \theta$ from $-\pi$ to $\pi$. The channel parameters are $(T,R_1,R_2,\theta) = (0.9,0.03,0.03,1)$. (b) The rate regions (blue dot and lines) and time-sharing rate regions (grey lines) under thermal noise $N_{\text{th}}=2$. The four subfigures are at $T=0.9$ with different reflection coefficients $(R_1,R_2)=(0.03,0.03),(0.03,0),(0.003,0.003),(0,0)$ respectively. (c) The coherent information $I_1$ as a function of $N_1$ for thermal noise $N_{\text{th}}=2$ and vacuum noise $N_{\text{th}}=0$, where $N_2=0$.}
     \label{SI_fig1}
\end{figure}

\subsection{Thermal states vs general Gaussian states}
\label{SI_A1}
Let $\mathsf{R}_{\text{thermal}}$ and $\mathsf{R}_{\text{gaussian}}$ be the rate regions achieved with thermal states and general Gaussian states respectively. Obviously $\mathsf{R}_{\text{thermal}} \subseteq \mathsf{R}_{\text{gaussian}}$.
Below we prove that thermal states give rise to local optima of $\mathsf{R}_{\text{gaussian}}$. Here we assume fixed channel parameters $(T,R_1,R_2,\theta)$ and the rate regions are obtained by varying the signal encodings.

Consider a mapping $\bm{x} \rightarrow (I_1,I_2)$, where $\bm{x} \in \mathbb{R}^n$ parametrizes the input states. A point $\bm{x}_0$ is a local optimum if the two gradient vectors $\bm{v}_k = \partial I_k / \partial \bm{x}$ are in the opposite direction: $\bm{v}_1 = \lambda \bm{v}_2$ with $\lambda<0$.
For any small deviation $d \bm{x}$ around $\bm{x}_0$, the changes $d I_k = \bm{v}_k \cdot d\bm{x}$ satisfy $d I_1 = \lambda d I_2$. Therefore if $I_1$ increases $I_2$ must decrease, and vice versa. Since $I_1$ and $I_2$ cannot be increased at the same time, $\bm{x}_0$ is a local optimum.
Notice that local optimum requires $\lambda<0$, which is stronger than the low-rank Jacobian condition since the low-rank Jacobian condition also includes $\lambda \geq 0$.

For general Gaussian states $\bm{x} \equiv (N_1,N_2,r_1,r_2,\Delta \theta)$, where $(N_k,r_k,\theta_k)$ are from the decomposition of $\bm{\mathrm{V}}_k$ (Eq.~(\ref{eq:gaussian_decomposition})). Notice that $(I_1,I_2)$ only depend on the relative angle $\Delta \theta = \theta_1 - \theta_2$ between the two squeezing axes.
This leads to the symmetry $I_k(N_1,N_2,r_1,r_2,\Delta \theta) = I_k(N_1,N_2,-r_1,-r_2,\Delta \theta)$ for $k=1,2$, since $r_1 \rightarrow -r_1$ and $r_2 \rightarrow -r_2$ together is equivalent to rotating the squeezing axis by 90 degree which keeps $\Delta \theta$ unchanged.
From this symmetry, we have
\begin{equation}
    \frac{\partial I_k}{\partial r_1} = \frac{\partial I_k}{\partial r_2} = \frac{\partial I_k}{\partial \Delta\theta} = 0 ,
\end{equation}
at $r_1=r_2=0$ for any $(N_1,N_2)$.

For thermal states $(N_1,N_2)$ at the boundary $\partial \mathsf{R}_{\text{thermal}}$, the gradient vectors must satisfy $(\partial I_1 / \partial N_1, \partial I_1 / \partial N_2) = \lambda (\partial I_2 / \partial N_1, \partial I_2 / \partial N_2)$ with $\lambda < 0$. These thermal states correspond to general Gaussian states at $\bm{x}_0=(N_1,N_2,0,0,0)$, where the gradient vectors $\bm{v}_k = (\partial I_k / \partial N_1, \partial I_k / \partial N_2, 0,0,0)$ also satisfy $\bm{v}_1 = \lambda \bm{v}_2$ with $\lambda<0$. This proves that thermal states are the local optima of $\mathsf{R}_{\text{gaussian}}$.

We numerically calculate $\mathsf{R}_{\text{gaussian}}$ by sweeping $(N_1,N_2,r_1,r_2,\Delta \theta)$ (Fig.~\ref{SI_fig1}(a) blue region), and compare it with $\partial \mathsf{R}_{\text{thermal}}$ (Fig.~\ref{SI_fig1}(a) red line and red dots). The result gives $\mathsf{R}_{\text{thermal}} = \mathsf{R}_{\text{gaussian}}$, which indicates that thermal states are likely the global optima among general Gaussian states.
The same conclusion also holds for other channel parameters $(T,R_1,R_2,\theta)$ that we tested.

\subsection{Thermal noise from the internal loss channels}
\label{SI_A2}
So far we have only considered vacuum noise injected from the internal loss channels, while thermal noises are common in practical devices, such as the pump-induced heating in microwave-optical quantum transducers~\cite{han2021}.
Here we take into account the effects of thermal noises by assuming all internal loss channels have $N_{\text{th}}$ average thermal occupation, i.e., $\bm{\mathrm{V}}_{n>2}= (\frac{1}{2} + N_{\text{th}}) \bm{\mathrm{I}}$.
The more general situations where the internal loss channels have different thermal occupations can be studied similarly.
The achievable rates under thermal noises are also given by Eq.~(\ref{eq:rates}), except now the output photon numbers are
\begin{equation}
    \begin{split}
        N_1' =& T N_1+R_2 N_2 + (1-T-R_2) N_{\text{th}} \\
        N_2' =& T N_2 + R_1 N_1 + (1-T-R_1) N_{\text{th}} .
    \end{split}
\end{equation}

We calculate the rate regions for $N_{\text{th}}=2$ in Fig.~\ref{SI_fig2}(b) at different transmission and reflection coefficients. The regions are much smaller than the ones with vacuum noise $N_{\text{th}}=0$ (main text Fig.~\ref{fig2}(b)).
The highest rates achievable with thermal states (Fig.~\ref{SI_fig2}(b) blue dots) are given by~\cite{holevo2001}
\begin{equation}
    I_{k,\text{max}} = \log_2 \frac{T}{1-T} - h (\bar{N}_k)
\end{equation}
where $\bar{N}_1 = (1-T-R_2) N_{\text{th}} /(1-T)$ and $\bar{N}_2 = (1-T-R_1) N_{\text{th}} /(1-T)$.

With thermal noise, one-side reflectionless condition no longer leads to vanishing discontinuity of $\partial \mathsf{R}$, which is different from the vacuum noise case. In the second subfigure of Fig.~\ref{SI_fig2}(b), we have $R_2=0$ while $\partial \mathsf{R}$ is still discontinuous at the $I_2$ axis (see inset).
This is because to achieve the highest rate $I_{2,\text{max}}$ on the $I_2$ axis, we need to have minimal added thermal noise which requires $N_1=0$. On the other hand, to achieve $I_1>0$ under thermal noise $N_{\text{th}}=2$, $N_1$ must exceed some threshold that is larger than 0 (Fig.~\ref{SI_fig2}(c) blue dot).

\section{Different settings of duplex quantum transduction}
\label{SI_B}
In this section, we discuss a few different settings to operate a quantum transducer in the duplex scenario, based on the number of users involved and whether or not the transduction is assisted with LOCC.

\begin{figure}
    \centering
    \includegraphics[width=0.48\textwidth]{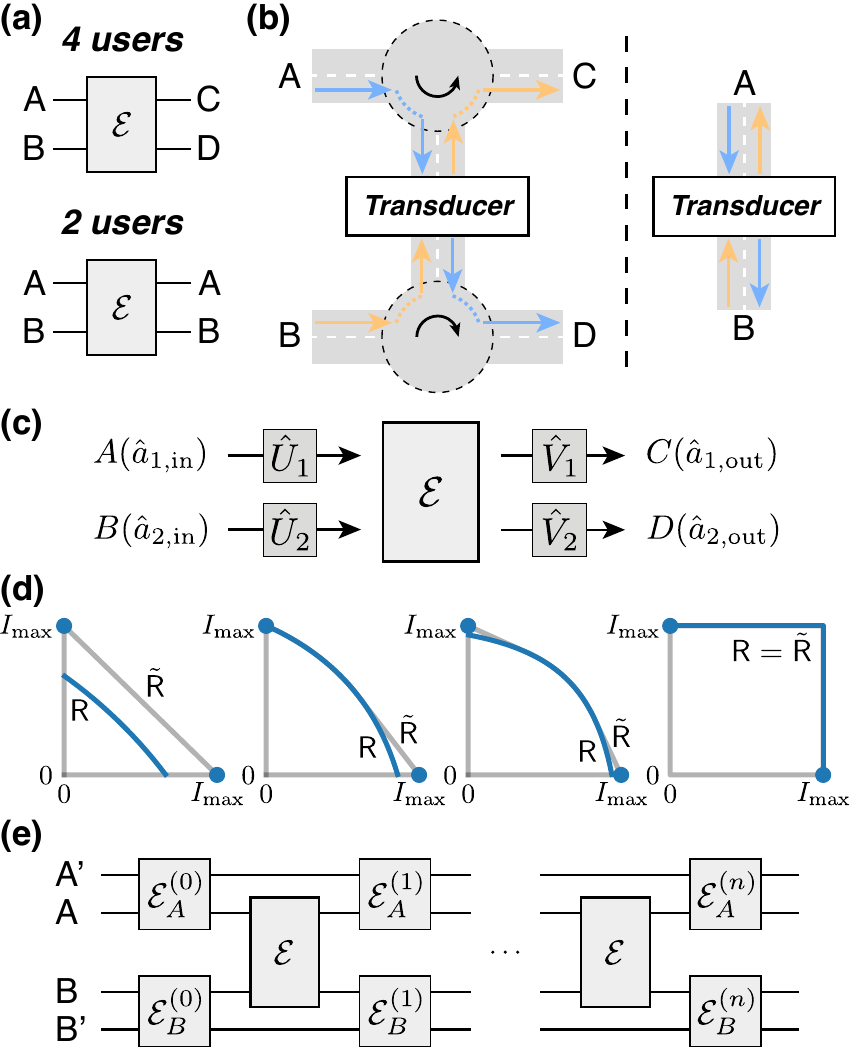}
    \caption{(a) Different number of users participating in duplex quantum transduction. 4 users: senders $A$ and $B$, receivers $D$ and $C$. 2 users: senders $A$ and $B$, receivers $B$ and $A$. (b) Schematic device layout for the 4-user setting (left) and the 2-user setting (right). In the 4-user setting, the input signals $A$ and $B$ are isolated from the output signals $C$ and $C$ with two circulators. (c) Invariance of the capacity region under local unitaries for simultaneous duplex transduction. (d) The rate regions (blue dot and lines) and time-sharing rate regions (grey lines) for LOCC-assisted duplex transduction at $T=0.9$ for different reflection coefficients $(R_1,R_2)=(0.03,0.03),(0.03,0),(0.003,0.003),(0,0)$. (e) In the 2-user setting, local operations $\mathcal{E}_{A(B)}^{(k)}$ are possible between uses of the bidirectional quantum channel $\mathcal{E}_{(A,B)\rightarrow (A,B)}$, which may also involve local ancilla $A'$ and $B'$.}
    \label{SI_fig2}
\end{figure}

\textbf{4 users, no LOCC.} In the main text, we focus on the 4-user setting, where the senders $A$ and $B$ are different from the receivers $C$ and $D$ (Fig.~\ref{SI_fig2}(a)). The 4-user setting can be realized by separating the input and output signal of each mode with a circulator (Fig.~\ref{SI_fig2}(b)). In this case, we could formally define the capacity region and study the symmetry of the capacity region.

The quantum capacity of a quantum channel is the maximal coherent information over all possible input states, including entangled states across multiple channel uses~\cite{schumacher1996,lloyd1997,devetak2005a}.
We can generalize quantum capacity to the capacity region for simultaneous duplex transduction. For a quantum interference channel $\mathcal{E}_{(A,B)\rightarrow (C,D)}: \mathcal{L}(\mathcal{H}_1 \otimes \mathcal{H}_2) \rightarrow \mathcal{L}(\mathcal{H}_1 \otimes \mathcal{H}_2)$, the capacity region is given by
\begin{equation}
    \mathsf{Q} (\mathcal{E}) \equiv \lim_{n\rightarrow \infty} \left\{\left(\frac{1}{n} I_1 \left(\mathcal{E}^{\otimes n}, \hat{\rho}^{(n)} \right), \frac{1}{n} I_2 \left(\mathcal{E}^{\otimes n}, \hat{\rho}^{(n)} \right) \right) \right\} 
\end{equation}
for all $\hat{\rho}^{(n)} = \hat{\rho}_1^{(n)} \otimes \hat{\rho}_2^{(n)}$ with $\hat{\rho}_k^{(n)} \in \mathcal{D} \left(\mathcal{H}_k^{\otimes n} \right), k=1,2$. Here $\mathcal{D}(\mathcal{H})$ is the space of density matrices on $\mathcal{H}$ and $I_1,I_2$ are defined in Eq.~(\ref{eq:general_rates}).

Since the capacity region is invariant under arbitrary single-mode unitaries before and after the quantum channel (Fig.~\ref{SI_fig2}(c)), we can exploit this symmetry to obtain equivalent channels with the same capacity region.
More specifically, we perform single-mode rotations which gives
\begin{equation}
    \bm{\mathrm{S}} \xLeftrightarrow{\mathsf{Q}} \bm{\mathrm{S}}_V \bm{\mathrm{S}} \bm{\mathrm{S}}_U ,
\end{equation}
where
\begin{equation}
    \begin{split}
        \bm{\mathrm{S}}_U =& \text{diag} (\bm{\mathrm{R}} (\alpha_1), \bm{\mathrm{R}} (\alpha_2), \bm{\mathrm{I}}, \cdots) \\
        \bm{\mathrm{S}}_V =& \text{diag} (\bm{\mathrm{R}} (\beta_1), \bm{\mathrm{R}} (\beta_2), \bm{\mathrm{I}}, \cdots) .
    \end{split}
\end{equation}
We choose the rotation angles $\alpha_1=0, \alpha_2=\theta_{11}-\theta_{12}, \beta_1=-\theta_{11},\beta_2 = -\theta_{21}$ and the symplectic matrix becomes
\begin{equation}
    \bm{\mathrm{S}}_V \bm{\mathrm{S}} \bm{\mathrm{S}}_U = 
    \begin{pmatrix}
        \sqrt{R_1} \bm{\mathrm{I}} & \sqrt{T_{12}} \bm{\mathrm{I}} & \cdots \\
        \sqrt{T_{21}} \bm{\mathrm{I}} & \sqrt{R_2} \bm{\mathrm{R}}(\theta) & \cdots \\
        \vdots & \vdots & \ddots
    \end{pmatrix} ,
\end{equation}
where $\theta = \theta_{11}+\theta_{22} - \theta_{12} - \theta_{21}$. Therefore, a scattering matrix $S$ has the same capacity region as a $2\times 2$ scattering matrix for the signal ports
\begin{equation}
    S \xLeftrightarrow{\mathsf{Q}} \begin{pmatrix}
        \sqrt{R_1} & \sqrt{T_{12}} \\
        \sqrt{T_{21}} & \sqrt{R_2} e^{i\theta}
    \end{pmatrix} .
\end{equation}

\textbf{4 users, with LOCC.} We consider duplex quantum transduction in the 4-user setting where each transduction channel is assisted with LOCC. For a quantum channel $\mathcal{E}: \mathcal{L}(\mathcal{H}) \rightarrow \mathcal{L}(\mathcal{H})$, the \emph{LOCC-assisted capacity}~\footnote{LOCC-assisted capacity is also known as ``two-way capacity'', but we avoid the term ``two-way'' to prevent any confusion with ``duplex'' quantum transduction.} is the maximal quantum information rate when the forward quantum channel is assisted with forward and backward classical communication~\cite{bennett1997}.
The LOCC-assisted capacity is lower bounded by the reverse coherent information~\cite{garcia-patron2009}
\begin{equation}
    \label{eq:reverse_coherent_information}
    I_R(\mathcal{E}, \hat{\rho}) = H(\hat{\rho}) - H ((\mathcal{E}\otimes\mathcal{I}') (\ket{\psi} \bra{\psi})), 
\end{equation}
where $\ket{\psi}$ is a purification of $\hat{\rho}$.
For a quantum interference channel $\mathcal{E}_{(A,B)\rightarrow (C,D)}: \mathcal{L}(\mathcal{H}_1 \otimes \mathcal{H}_2) \rightarrow \mathcal{L}(\mathcal{H}_1 \otimes \mathcal{H}_2)$, the LOCC assisted achievable rates $(I_1,I_2)$ with input state $\hat{\rho}_1 \otimes \hat{\rho}_2$ are
\begin{equation}
    \begin{split}
        I_1 \left(\mathcal{E}, \hat{\rho}_1 \otimes \hat{\rho}_2 \right) \equiv& I_R \left(\mathcal{E}_1, \hat{\rho}_1 \right) \qquad \mathcal{E}_1 (\cdot) = \tr_1 \mathcal{E} (\cdot, \hat{\rho}_2) \\
        I_2 \left(\mathcal{E}, \hat{\rho}_1 \otimes \hat{\rho}_2 \right) \equiv& I_R \left(\mathcal{E}_2, \hat{\rho}_2 \right) \qquad \mathcal{E}_2 (\cdot) = \tr_2 \mathcal{E} (\hat{\rho}_1, \cdot) .
    \end{split}
\end{equation}
and negative $I_{1(2)}$ are set to 0.

We consider linear transducers with thermal input states $(N_1,N_2)$. The achievable rates are
\begin{equation}
    \begin{split}
        & I_k (T,R_1,R_2,N_1,N_2) = h(N_k) \\
        & - h\left( \frac{D_k + N_k' - N_k - 1}{2} \right) - h\left( \frac{D_k - N_k' + N_k - 1}{2} \right) .
    \end{split}
\end{equation}
The rate region $\{(I_1,I_2)| \forall (N_1,N_2)\}$ gives a lower bound on the LOCC-assisted capacity region.

In Fig.~\ref{SI_fig2}(d), we plot the rate regions (blue dots and lines) as well as their convex hulls (grey lines) for different reflection coefficients. Here $T=0.9$ and $I_{\text{max}} = -\log_2 (1-T) \approx 3.32$ is the LOCC-assisted capacity for pure-loss channel~\cite{pirandola2017}.

\textbf{2 users.} Alternatively, we may also consider the 2-user (bipartite) setting where $A$ and $B$ are both the senders and receivers (Fig.~\ref{SI_fig2}(a)). The 2-user duplex quantum transduction can be modeled as a bidirectional quantum channel $\mathcal{E}_{(A,B)\rightarrow (A,B)}$~\cite{das2021}. The key difference from the 4-user setting is that local operations are possible between different uses of the transducer (Fig.~\ref{SI_fig2}(e)). 
Without classical communication between $A$ and $B$~\cite{bennett2003,childs2006,lau2019,zhang2022a,ding2023}, the performance of the transducer can still be characterized by the capacity region $\{(I_{1,A\rightarrow B}, I_{2,B\rightarrow A}) \}$.
With classical communication, however, the entanglement generation capacity becomes the performance metric instead of the capacity region, since $A$ and $B$ can perform time-shared quantum teleportation using entanglement generated during the channel uses.
Upper bounds on the entanglement capacity of bidirectional channels have been studied before~\cite{bauml2018,das2020,das2021}.

\section{Optimality of two-side reflectionless condition}
\label{SI_C}
Here we prove that two-side reflectionless condition leads to optimal duplex quantum transduction within a more general setting.
For linear transducers that we considered, a larger transmission coefficient with smaller reflection coefficients strictly leads to a better rate region.
In other words, $\mathsf{R}(T,R_1,R_2) \supseteq \mathsf{R}(T',R_1',R_2')$ if $T\geq T',R_1\leq R_1', R_2 \leq R_2'$.
Below we show that two-side reflectionless condition gives the largest transmission and therefore the optimal rate region.

Consider a general $N$-mode linear Hamiltonian $\oph = \sum G_{mn} \opad_m \opa_n$, where $G=G^\dagger$ and $G_{kk} \equiv \Delta_k$ is the detuning of mode $k$ in the rotating frame.
Mode $k$ also has external (internal) loss rate $\kappa_{k,e(i)}$. 
The input-output relation is given by
\begin{equation}
    \begin{pmatrix}
        \opa_{1,\text{out}} \\
        \vdots \\
        \opa_{N,\text{out}} \\
        \tilde{a}_{1,\text{out}} \\
        \vdots \\
        \tilde{a}_{N,\text{out}} 
    \end{pmatrix} = S
    \begin{pmatrix}
        \opa_{1,\text{in}} \\
        \vdots \\
        \opa_{N,\text{in}} \\
        \tilde{a}_{1,\text{in}} \\
        \vdots \\
        \tilde{a}_{N,\text{in}} 
    \end{pmatrix} ,
\end{equation}
where $\opa_{k,\text{in(out)}}$ is the input (output) operator for the external coupling of mode $k$ and $\tilde{a}_{k,\text{in(out)}}$ is the input (output) operator for the internal loss of mode $k$.
The scattering matrix is
\begin{equation}
    \label{eq:general_S}
    S = 
    \begin{pmatrix}
    I+\sqrt{K_e} M \sqrt{K_e} & \sqrt{K_e} M \sqrt{K_i} \\
    \sqrt{K_i} M \sqrt{K_e} & I+\sqrt{K_i} M \sqrt{K_i}
    \end{pmatrix} ,
\end{equation}
where $I$ is the $n\times n$ identity matrix and
\begin{equation}
    \begin{split}
        M =& -(iG + (K_{e} + K_{i})/2)^{-1} \\
        K_{e(i)}=& \text{diag}(\kappa_{1,e(i)},...,\kappa_{N,e(i)}) .
    \end{split}
\end{equation}

We only consider external coupling ports with $S=I+\sqrt{K_e} M \sqrt{K_e}$.
The power transmission coefficient from port $n$ to port $m$ is $T_{mn}=|S_{mn}|^2$, where $S_{mn}=\sqrt{\kappa_{m,e}\kappa_{n,e}} M_{mn}$ for $m \neq n$.
The power reflection coefficient for port $n$ is $R_n = |S_{nn}|^2$. We would like to find conditions that maximize the transmission.

Notice that for any variable $\alpha$
\begin{equation}
    0 = \frac{\partial \left( M M^{-1} \right)}{\partial \alpha} \quad \Rightarrow \quad \frac{\partial M}{\partial \alpha} = - M \frac{\partial M^{-1}}{\partial \alpha} M ,
\end{equation}
which leads to
\begin{equation}
    \begin{split}
        \frac{\partial M_{mn}}{\partial \kappa_{k,e}} =& \frac{1}{2} M_{mk} M_{kn} \\
        \frac{\partial M_{mn}}{\partial \Delta_k} =& i M_{mk} M_{kn} ,
    \end{split}
\end{equation}
as well as
\begin{equation}
    \begin{split}
        \frac{\partial S_{mn}}{\partial \kappa_{k,e}} =& \frac{1}{2\kappa_{k,e}} S_{mk} S_{kn} \qquad \forall m\neq n \\
        \frac{\partial S_{mn}}{\partial \Delta_k} =& = \frac{i}{\kappa_{k,e}} S_{mk} S_{kn} \qquad \forall m, n .
    \end{split}
\end{equation}
Therefore we have
\begin{equation}
    \begin{split}
        \frac{\partial T_{mn}}{\partial \kappa_{n,e}} =& \frac{T_{mn}}{\kappa_{n,e}} \text{Re} \{S_{nn}\} \\
        \frac{\partial T_{mn}}{\partial \kappa_{m,e}} =& \frac{T_{mn}}{\kappa_{m,e}} \text{Re} \{S_{mm}\} ,
    \end{split}
\end{equation}
and
\begin{equation}
    \begin{split}
        \frac{\partial T_{mn}}{\partial \Delta_n} =& -\frac{2 T_{mn}}{\kappa_{n,e}} \text{Im} \{S_{nn}\} \\
        \frac{\partial T_{mn}}{\partial \Delta_m} =& -\frac{2 T_{mn}}{\kappa_{m,e}} \text{Im} \{S_{mm}\} .
    \end{split}
\end{equation}
We have reached the conclusion:
if the detuning $\Delta_{m(n)}$ and external coupling rate $\kappa_{m(n),e}$ are tunable for mode $m(n)$, then maximizing $T_{mn}$ leads to zero reflection from mode $m(n)$ since
\begin{equation}
    \frac{\partial T_{mn}}{\partial \Delta_{m(n)}} = \frac{\partial T_{mn}}{\partial \kappa_{m(n),e}} = 0 \quad \Rightarrow \quad S_{mm(nn)} =0 .
\end{equation}
If the detunings and external coupling rates for both ports $m$ and $n$ are tunable, maximizing $T_{mn}$ gives the two-side reflectionless condition $R_m=R_n=0$.

Back to the two-mode case where $\Delta_1,\Delta_2,\kappa_{1,e},\kappa_{2,e}$ are tunable. Solving the necessary condition $R_1=R_2=0$ leads to a unique solution of
$\Delta_1=\Delta_2=0$ as well as
\begin{equation}
    \label{eq:optimal_solution}
    \begin{split}
         \kappa_{1,e}^2 =& \frac{\kappa_{1,i}}{\kappa_{2,i}} (4g^2 + \kappa_{1,i} \kappa_{2,i}) \\
         \kappa_{2,e}^2 =& \frac{\kappa_{2,i}}{\kappa_{1,i}} (4g^2 + \kappa_{1,i} \kappa_{2,i}) .
    \end{split}
\end{equation}
Notice that $T$ takes its maximum at finite values of $\Delta_1,\Delta_2,\kappa_{1,e},\kappa_{2,e}$. Due to the uniqueness of the solution, two-side reflectionless condition must correspond to the global maximum of the transmission coefficient, and thus is optimal for duplex quantum transduction.

Furthermore, in the two-side reflectionless case, the two channels $\mathcal{E}_{1,A\rightarrow D}$ and $\mathcal{E}_{2,B\rightarrow C}$ are completely decoupled. This is true not just for thermal input states, but also for general quantum states $\hat{\rho}_1,\hat{\rho}_2$.
In other words, $I_1 \left(\mathcal{E}, \hat{\rho}_1 \otimes \hat{\rho}_2 \right)$ does not depend on $\hat{\rho}_2$ and $I_2 \left(\mathcal{E}, \hat{\rho}_1 \otimes \hat{\rho}_2 \right)$ does not depend on $\hat{\rho}_1$.

\begin{figure}
    \centering
    \includegraphics[width=0.48\textwidth]{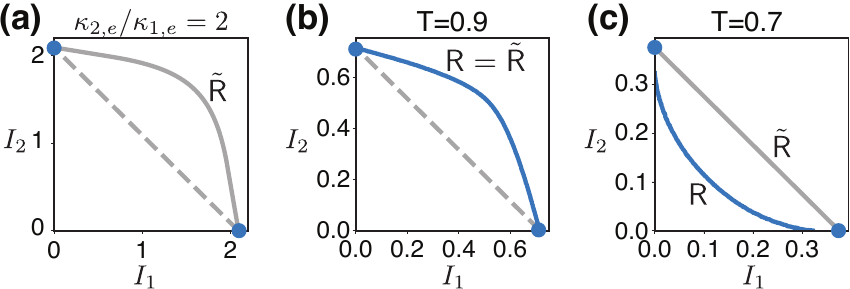}
    \caption{(a) Time-sharing rate region for asymmetric loss rates $\kappa_{2,e}/\kappa_{1,e}=2$ with $\sqrt{\kappa_{1,e} \kappa_{2,e}}=10$. Here $g=5, \kappa_{1,i} = \kappa_{2,i} = 1$ and we optimize over the detunings of the transducer. (b-c) Rate regions achieved with the $\{ \ket{0}, \ket{1}\}$ encodings for the lossless beam splitter channel at $T=0.9$ and $T=0.7$.}
     \label{SI_fig3}
\end{figure}

\section{Two-mode transducer with asymmetric loss rates}
\label{SI_D}
In the main text, we studied two-mode quantum transducer with symmetric loss rates. Here we generalize the discussion to asymmetric loss rates.
The scattering matrix now is given by Eq.~(\ref{eq:general_S}) for $N=2$, instead of Eq.~(\ref{eq:S_matrix}) in the main text. Notice that $S$ only depends on the ratios between the device parameters since it is invariant under the transformation
\begin{equation}
    \begin{split}
        \kappa_{k,i} \rightarrow & 1 \\
        \kappa_{k,e} \rightarrow & \kappa_{k,e}/\kappa_{k,i} \\
        G \rightarrow & (K_i)^{-1/2} G (K_i)^{-1/2} .
    \end{split}
\end{equation}
For a two-mode transducer, $S$ only depends on $\kappa_{k,e}/\kappa_{k,i}$, $\Delta_k/\kappa_{k,i}$ for $k=1,2$ and $g/\sqrt{\kappa_{1,i} \kappa_{2,i}}$. Therefore we set the internal loss rates of both modes to be 1 without loss of generality, and only study the ratios between the external and internal loss rates.
In practice, the loss rates of two modes may differ by orders of magnitude, but the ratios $\kappa_{1,e}/\kappa_{1,i}$ and $\kappa_{2,e}/\kappa_{2,i}$ are often close~\cite{fan2018,mckenna2020,xu2021}.

In Fig.~\ref{SI_fig3}(a), we plot the time-sharing rate region for the asymmetric case $\kappa_{2,e}/\kappa_{1,e}=2$ with $\sqrt{\kappa_{1,e} \kappa_{2,e}}=10$, which demonstrates the benefit from the simultaneous duplex transduction with asymmetric loss ratios.
The optimal rate region achieved with the two-side reflectionless condition Eq.~(\ref{eq:optimal_solution}) requires symmetric ratios $\kappa_{1,e}/\kappa_{1,i} = \kappa_{2,e}/\kappa_{2,i}$.
Therefore the rate regions with asymmetric loss ratios (Fig.~\ref{SI_fig3}(a)) are all subsets of the optimal rate region in Fig.~\ref{fig3}(b)iv.

\section{$\{ \ket{0}, \ket{1}\}$ encodings}
\label{SI_E}
Although we focus on Gaussian encodings throughout the paper, non-Gaussian states such as single photon states may be easier to implement with current technologies.
Here we restrict to the simplest non-Gaussian encodings, the $\{ \ket{0}, \ket{1}\}$ encodings, where the rate regions are obtained from the coherent information (Eq.~(\ref{eq:general_rates})) for all density matrices in the 0 and 1 photon subspace, i.e., $\hat{\rho}_1,\hat{\rho}_2 \in \mathcal{D}(\{ \ket{0}, \ket{1}\})$.

We calculate the rate regions for the lossless beam splitter channel (Eq.~(\ref{eq:beam_splitter})) at different transmission coefficients. The simultaneous duplex transduction outperforms the time-shared unidirectional transduction for large enough $T$ (Fig.~\ref{SI_fig3}(b)).
This makes sense since the lossless beam splitter approaches the ideal SWAP operation as $T\rightarrow 1$ with a square rate region.
Interestingly, $\partial \mathsf{R}$ is discontinuous at both $I_1,I_2$ axes for $T=0.7$ (Fig.~\ref{SI_fig3}(c)) while we do not observe any discontinuity at $T=0.9$, although both cases have non-zero reflection coefficients $1-T$.

\bibliography{Two_way_transduction.bib}

\end{document}